\documentclass[useAMS,usenatbib]{mn2e}
\usepackage{epsfig,graphicx,latexsym,amsmath,amssymb}
\usepackage{natbib}
\usepackage{hyperref}
\usepackage{mathrsfs}
\usepackage{url}

\bibpunct[,]{(}{)}{;}{a}{}{,}

\title[WDs as a probe to DM crests]
      {Probing dark matter crests with white dwarfs and IMBHs}

\author[Pau Amaro-Seoane et al.]
{P. Amaro-Seoane$^{1}$\thanks{E-mail: Pau.Amaro-Seoane@aei.mpg.de (PAS)},
J. Casanellas$^{1}$, R. Sch{\"o}del$^2$, E. Davidson$^{1}$ \& J. Cuadra$^{3}$
   \\
$^{1}$Max Planck Institut f\"ur Gravitationsphysik
(Albert-Einstein-Institut), D-14476 Potsdam, Germany\\
$^{2}$Instituto de Astrof\'isica de Andaluc\'ia (CSIC), Glorieta de la Astronom\'ia s/n, 18008 Granada, Spain\\
$^{3}$Instituto de Astrof\'isica, Pontificia Universidad Cat\'olica de Chile, Santiago, Chile\\
}

\begin{document}

\date{draft \today}

\pagerange{\pageref{firstpage}--\pageref{lastpage}} \pubyear{2015}

\maketitle

\label{firstpage}

\begin{abstract}
White dwarfs (WDs) are the most promising captors of dark matter (DM) particles
in the crests that are expected to build up in the cores of dense stellar
clusters. The DM particles could reach sufficient densities in WD cores to liberate energy through self-annihilation. The extinction associated with our Galactic Centre, the most promising region where to look for such effects, makes it
impossible to detect the potential associated luminosity of the DM-burning WDs because due to distance and extreme extinction the apparent near-infrared magnitudes of the WDs would be fainter than about 30 mag.
However, in smaller stellar systems which are close enough to us and not
heavily extincted, such as $\omega-$Cen, we may be able to detect DM-burning WDs. In
this work we investigate the prospects of detection of DM-burning WDs in a
stellar cluster harbouring an intermediate-mass black hole (IMBH), which leads
to higher densities of DM at the centre as compared with clusters without one.
We calculate the capture rate of WIMPs by a WD around an IMBH and estimate the
luminosity that a WD would emit depending on its distance to the center of the
cluster.  Direct-summation $N-$body simulations of $\omega-$Cen yield a
non-negligible number of WDs in the range of radii of interest.
We apply our assumption to published HST/ACS observations of stars in the center of $\omega-$Cen
to search for DM burning WDs and, although we are not able to identify any
evident candidate because of crowding and incompleteness, we proof that
their bunching up at high
luminosities would be unique.
We predict that DM burning
will lead to a truncation of the cooling sequence at the faint
end.
The detection of DM burning in future observations of dense stellar
clusters, such as globular clusters or ultra-compact dwarf galaxies could allow
us to probe different models of DM distributions and characteristics such as
the DM particle scattering cross section on nucleons. On the other hand, if DM-burning WDs really exist, their number and properties could give hints to the existence of IMBHs.
\end{abstract}


\section{Introduction}
\label{sec.intro}

Weakly interacting massive particles (WIMPs) form high density cusps in dense
stellar systems, as seen with collisionless cosmological $N-$body simulations
\citep{NavarroFrenkWhite97,MooreEtAl99}. More recent simulations have derived
the result that this is particularly true in the gravitational well of massive
black holes \citep[henceforth MBHs;
see][]{GondoloSilk99,GnedinPrimack2004,BertoneMerritt05}.  Within a certain
radius of the MBH any member of the stellar distribution has a big WIMP capture
rate. The successive annihilation of these particles in the core of the stars
releases a significant amount of energy and hence impinges their evolution and
appearance
\citep{SalatiSilk89,BouquetSalati89,MoskalenkoWai07,Scott:2008ns,Casanellas:2009dp}.
In the case of main-sequence stars, the first two references describe the
potential suppression of stellar core convection in these stars, so that there
could be an agglomeration of stars in our Galactic Center concealing their
spectral type and being interpreted as cold red giants. \cite{MoskalenkoWai07}
found that the WIMP capture rate and annihilation is remarkably large for WDs.
They derived luminosities from the WIMP annihilation only that are comparable to
or even larger than the standard thermal luminosity of WDs, of the order
$L_{\rm WD}\sim 3\times10^{34}$ erg s$^{-1} \sim 10\,L_{\odot}$, with
$L_{\odot}$ the luminosity of the sun. In this respect, WDs are the most
promising candidates to probe DM agglomeration zones, such as our galactic
nucleus or the center of dwarf spheroidal galaxies and globular
clusters~\citep{2008PhRvD..77d3515B,Hooper:2010es,McCullough:2010ai,2011PhRvD..84j3510F,2014arXiv1410.3925H}.

This article is organised as follows: In section~\ref{sec.rhochi} we calculate the distribution of DM particles
around a massive black hole such as the one that might be lurking in
$\omega-$Cen.  Later, in section~\ref{sec.Lum} we estimate the capture
rates and, accordingly, the luminosities emitted by WDs of different masses and
compositions.  In
section~\ref{sec.DistribWDs} we model the distribution of WDs as a function of
radius in a cluster harbouring a massive black hole.
We discuss observational predictions and compare them with existing photometric data
in section~\ref{sec.Observ}.
Finally, we summarise and conclude the most important implications
of our analysis in section~\ref{sec.conc}.

\section{Dark matter density profile at $\omega-$Cen}
\label{sec.rhochi}

Although it is not possible to precisely determine the quantity of dark matter (DM)
present in globular clusters, we can estimate the DM density profile
in $\omega-$Cen based on current observations and simulations. These clusters are thought
to have formed in the center of DM subhaloes in the early Universe, thus being
born as strongly DM dominated objects~\citep{art-Peebles1984}. However, in the scenario of
hierarchical structure formation, clusters were captured by galaxies, losing most of
their extra-nuclear mass due to tidal stripping, but still retaining large
quantities of DM concentrated in their cores, as indicated by the results of
numerical simulations~\citep{Tsuchiya:2003ev,Mashchenko:2004hk,Ibata:2012eq}.

It is important to note that $\omega-$Cen may not be a normal globular cluster. It is
considered rather likely that it is the stripped nuclar cluster of a tidally
accreted galaxy which, reinforced by its far more complex stellar population,
compared to normal clusters.  The fact that $\omega-$Cen may be a nuclear star
cluster remnant is also used to argue that it may have an IMBH at its center
(see e.g. the introduction of \citealt{2010ApJ...710.1063V}).

In the case of $\omega-$Cen, with a cluster mass of
$2.5\times10^6\;$M$_{\odot}$~\citep{vandeVen:2005hk}, we can estimate the
initial mass of the DM subhalo to be $5\times10^8\;$M$_{\odot}$
($M_{DM,0}\approx0.0038^{-1} M_{GC}$, with $M_{GC}$ the mass of the globular cluster, see~\citealt{2010MNRAS.405..375G}), and its
present value to be $10^7\;$M$_{\odot}$ ($M_{DM}\approx 0.02 M_{DM,0}$,
see~\citealt{Gao:2004au}). The density of the initial DM halo in $\omega-$Cen was
modelled assuming a NFW profile~\citep{Navarro:1995iw}:

\begin{equation}
\rho_{\chi,\textmd{NFW}}(r)=\frac{\rho_c}{(r/r_s)(1+r/r_s)^2} ,
\end{equation}

\noindent
with $r_s=685\;$pc and $\rho_c=0.07\;$M$_{\odot}/$pc$^3$. This parametrization
of $\rho_{\chi}$ correctly reproduces the results of DM N-body simulations, and
predicts a $\rho\propto r^{-1}$ cusp in the centre of the DM halo. Other
parametrizations would imply either steeper profiles~\citep{Moore:1997sg} or
cored profiles~\citep{Burkert:1995yz}, so this uncertainty should be taken into
account when interpreting our results.

It has been shown that the inclusion of the baryonic feedback leads to an
adiabatic contraction of the DM halo~\citep{art-Blumenthal86}. This mechanism
was implemented using the baryonic mass profile observed in
$\omega-$Cen~\citep{art-Noyola08}, following the procedure described
in~\cite{Gnedin:2004cx}, leading to a contracted profile
$\rho_{\chi,\textmd{NFW-AC}}(r)$. Furthermore, the DM cusp created by the
adiabatic contraction is shallowed by the heating of the DM particles due to
the collisions with the stars~\citep{Merritt:2003qk}, creating a core of
constant density up to $r_{\rm DMh}$, the radius at which the two-body
relaxation time defined by the stars, $T_r(r)$, becomes greater than the age of
the cluster. In $\omega-$Cen the $r_{\rm DMh}$ was found to be of approximately
3.5~pc~\citep{vandeVen:2005hk}.

In the case of DM haloes with a central MBH, the central cusp in the stellar
density leads to an overdensity in the inner region of the final DM profile,
the so-called \textit{crest}, as shown by Fokker-Planck and direct $N-$body
integrations~\citep{2004PhRvL..93f1302G,Merritt:2006mt}.

The characteristic time for the growth of a stellar cusp and a DM crest is
approximately $0.5\,T_r(r_h)$, where $r_h$ is the gravitational influence
radius of the central MBH. After this time, the stars around the MBH form a
Bahcall-Wolf cusp: $\rho_{\star}(r) \propto r^{7/4}$~\citep[see
e.g.][]{Peebles72,Bahcall:1976aa,ASEtAl04}, triggering the formation of a DM
crest with a density profile $\rho_{\chi}\propto r^{-1.5}$.  In $\omega-$Cen
there is no clear \textit{observed} evidence for the cusp, probably due to the
incompleteness of the star counts in the centre (crowding, see section
\ref{sec.Observ} on the limitations of the observations).  In this work we  we
assume the existence of the cusp, as expected theoretically in the works we
just mentioned.

Thus, the presence of an IMBH in the centre of $\omega-$Cen strongly impacts the
estimation of the DM distribution around it. However, the observational
evidence for its existence remains controversial. \cite{2010ApJ...719L..60N}
found evidence for an IMBH, but this is contested by
\cite{2010ApJ...710.1063V}. Key problems are the low angular resolution used in
previous spectroscopic work on the kinematics of stars and the determination of
the location of the center of the cluster. Consequently, here we will consider the
DM distribution in both scenarios, with and without an IMBH in the center of
$\omega-$Cen \citep{2010ApJ...710.1063V,2010ApJ...719L..60N}.

Asuming a central IMBH of mass $M_{\bullet}\approx 10^4\;$M$_{\odot}$,
$r_h$ was found to be equal to 0.12 pc (taking an heliocentric distance to
$\omega-$Cen of 4.8 kpc, so $r_h \sim 5'' \sim 0.12$ pc \citealt{JalaliEtAl2012}) and
$0.5T_r(r_h)\sim10^9\;$yr~\citep{vandeVen:2005hk}, well below the age of the
$\omega-$Cen. In this case, the final DM density profile of $\omega-$Cen was estimated to be:
\begin{equation}
\rho_{\chi, \bullet}(r)=\left\{
\begin{array}{ll}
\rho_{\chi,\textmd{core}} \times \left(r / r_h \right)^{-1.5}, & r<r_h,\\
\rho_{\chi,\textmd{core}} \equiv \rho_{\chi,\textmd{NFW-AC}}(r_{\rm DMh}), & r_h\leq r\leq r_{\rm DMh},\\
\rho_{\chi,\textmd{NFW-AC}}(r), & r>r_{\rm DMh},
\end{array}\right.
\end{equation}

\noindent
similar to what has been estimated in the literature for other clusters harbouring
IMBHs~\citep{art-HESS11,art-Feng12}.
Finally, we also consider an upper limit to the
DM density due to the DM annihilations,
$\rho_{ann} \approx m_{\chi} /
\langle\sigma_a v\rangle t_{\rm GC}$,
where $m_{\chi}$ is the WIMP mass, $\langle\sigma_a v\rangle$
its thermally averaged annihilation cross section, and $t_{\rm GC}$
the time since the formation of the crest, which we conservatively assumed
to be the age of $\omega-$Cen. This limit is approximately equal to $
2\times10^{7}$, $4\times10^{7}$ and $4\times10^{8}\;$M$_{\odot}/$pc$^3$
for WIMP masses of 5, 10 and 100 GeV.
These limits only flatten the DM density profile at distances below $10^{-3}$
pc.

\begin{figure}
\resizebox{\hsize}{!}
          {\includegraphics[scale=1,clip]{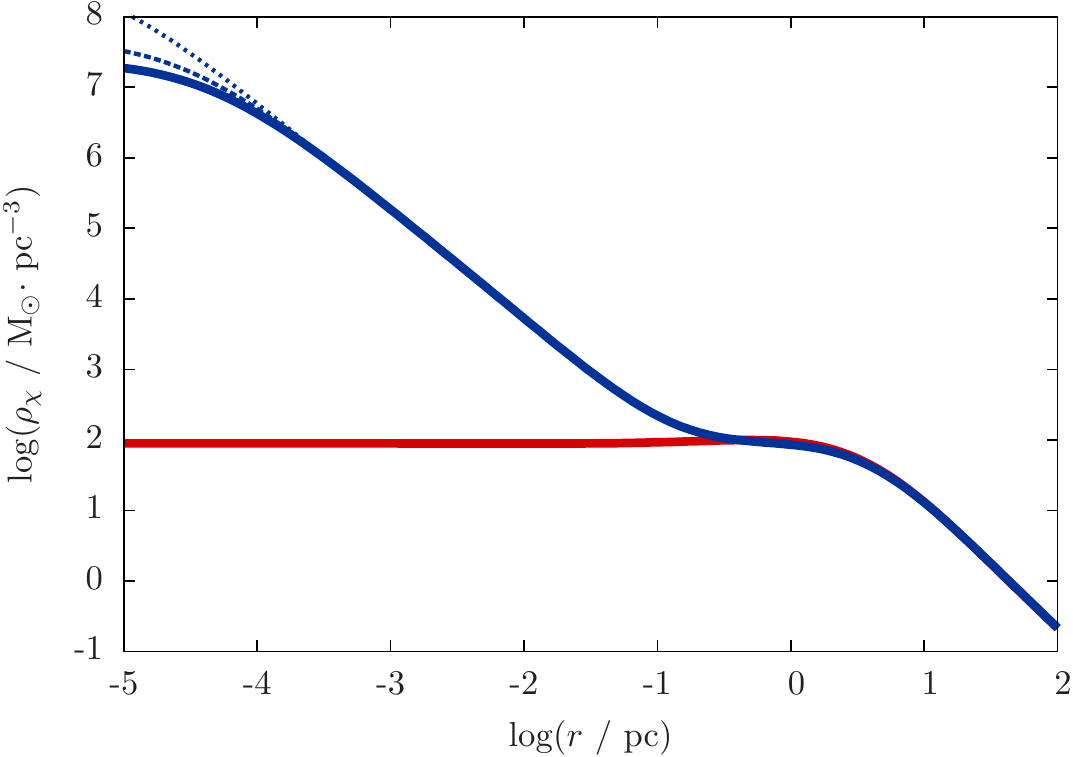}}
\caption
   {
Density profile of the DM halo in $\omega-$Cen. The red line shows $\rho_{\chi}(r)$
as if there was no IMBH in the center of the cluster, while the blue lines show $\rho_{\chi, \bullet}(r)$ for a IMBH of $M_{\bullet}\approx 10^4\;$M$_{\odot}$ for $m_{\chi}=$5 GeV (continuous), 10 GeV (dashed) and 100 GeV (dotted).
    }
\label{fig.rho}
\end{figure}

In the scenario where $\omega-$Cen harbours no IMBH, the DM distribution would remain cored in the center, leading to the the following DM density profile:
\begin{equation}
\rho_{\chi}(r)=\left\{
\begin{array}{ll}
\rho_{\chi,\textmd{core}} \equiv \rho_{\chi,\textmd{NFW-AC}}(r_{\rm DMh}), & r<r_{\rm DMh},\\
\rho_{\chi,\textmd{NFW-AC}}(r), & r\geq r_{\rm DMh},
\end{array}\right.
\end{equation} Both DM density profiles are shown in Figure~\ref{fig.rho}.

\section{Dark matter cusps and IMBHs}
\label{sec.Lum}

In order to calculate the number of WIMPs captured by the WD, we
have to calculate the capture rate ${\cal C}$ of a WD of mass $M_*$
for a Maxwellian WIMP velocity distribution. We assume the WD moving
at a velocity $v_*$ relative to the observer, so that the capture rate is \citep{Gould1987},

\begin{equation}
{\cal C}=4\pi \int_0^{R_*} dr\, r^2\, \frac{dC(r)}{dV},
\label{gould2.27}
\end{equation}

\noindent
where ${dC(r)}/{dV}$ can be calculated with the expressions of
the same work.
The total capture rate depends mainly on the density of DM around the
star, the density and composition of the WD, and the spin-independent
scattering cross-section of the DM particles off nucleons, $\sigma_{\chi}$.
 As noted by \cite{MoskalenkoWai07}, WD are the most
promising captors of WIMPs due to the fact that the cross section is
proportional to $A_n^4$, with $A_n$
the atomic number of the principal element of which the WD's nucleus is composed.
The maximum interaction cross-section $\sigma_{max}$ is given by the geometrical
limit $\pi R_*^2$, with $R_*$ the radius of the star, and can be calculated as
\begin{equation}
\sigma_{max} A_n^4 \frac{M_*}{M_n} = min (\sigma_{\chi} A_{n}{}^{4}
\frac{M_*}{M_n}, \pi R_*^2),
\end{equation}
where $M_n$ is the nucleus mass and $\sigma_{\chi}$ the cross section of the particle.
This leads to limits on our
cross section to values around $\sigma_{max} = 10^{-42}\;$cm$^2$, depending
on the composition of the WD and its mass.

We take the values of the distribution of $\rho_{\chi}$ obtained in section~\ref{sec.rhochi}
to estimate the capture rates.
The capture rate ${\cal C}$ can be easily converted into luminosities by using
$L_{\chi}={\cal C} \, m_{\chi}$ \citep{SalatiSilk89}.  The luminosities due to
DM annihilations for models of WDs with different compositions, masses and
radius are shown in Table~\ref{tab.all}.
The principal factor in the luminosity, and the related effective temperature, $T_{eff}$, is the radius of the star,
followed by the
atomic number. Oxygen WDs have the largest values, followed by Carbon
WDs and the lowest values are for Helium WDs. We display the value of the
luminosity due to the burning of DM at three particular radii taken from the
centre of the cluster, assumed to be located at the position of the IMBH:
$2.5$, $0.1$ and $0.01$ pc. While the DM density achieves higher values at
shorter radii, we choose $0.01$ pc as our lowest value, because
shorter radii will contain very small numbers of WDs, thus making any observational/statistical test inconclusive.
In the table we depict
the associated luminosities for the different kinds of WDs at different radii.
For all the cases studied here, $L_{\chi}$ does not depend on the mass of the
DM particle $m_{\chi}$ (because ${\cal C} \propto m_{\chi}^{-1}$ in this
regime).

We note that $m_{\chi}$ should be below 6 GeV to avoid the current limits
from direct detection experiments on the spin-independent DM-nucleon cross
sections~\citep{2014PhRvL.112i1303A}. However, given the present controversy
between contradictory positive and null results in different experiments,
it is worth to explore the DM parameter space more broadly. In particular,
for $m_{\chi}\approx10\;$GeV, our results show how WDs would be lightened
up if DM has the properties to explain the recent positive results in some
DM detectors~\citep{Bernabei:2008yi,2012EPJC...72.1971A,Agnese:2013rvf}.

\section{Distribution of white dwarfs in $\omega-$Cen}
\label{sec.DistribWDs}

An obvious question to address is that of the number of WDs available in the
range of radii of interest. $\omega-$Cen is a very massive globular cluster,
or a nuclear cluster remnant. Since we are interested in an accurate distribution of stars along the
radius, we use the results of a direct-summation $N-$body integration to model
this.

We employ the simulation data of \cite{McNamaraEtAl2012} with a 1\% MBH in
mass. We note that these simulations were for NGC 6266, but have been adapted to fit the profile of $\omega-$Cen. Ideally one would model the whole cluster with a direct-summation
integrator but the number of stars and the long integration time makes it
impossible. The simulation uses 1,580,430 stars, which is below the expected
number of stars for $\omega-$Cen (about 1/6 of the total number), but
representative and the relative distribution is correct, in the sense that that
model has the same half-mass radius as $\omega-$Cen. It hence represents a fair
lower-limit for the total number of WDs distributed along the radius for a
cluster as massive as $\omega-$Cen.

In the simulation we have three different
mass groups: light WDs, of masses $<\,0.6\,M_{\odot}$, medium ones, with masses
larger than $0.6\,M_{\odot}$ but less than $0.8\,M_{\odot}$, and heavy ones,
with masses $\leq\,0.8\,M_{\odot}$. For small radii, up to $0.01$ pc, we find in the simulation that there are
1 light WD, 0 medium ones, and 43 heavy ones. Between 0 and 0.5 pc we have 154
light WDs, 77 medium WDs and 95 heavy ones. Out to 1 pc there are 726 light
WDs, 255 medium ones and 189 heavy WDs. Up to 10 pc, 25378 light WDs, 4959 medium
ones, and 1864 heavy WDs.
We therefore find that the number of WDs at the radii of interest is non-negligible.

\section{Observational implications }
\label{sec.Observ}

\begin{figure}
\resizebox{\hsize}{!}
          {\includegraphics[scale=1,clip]{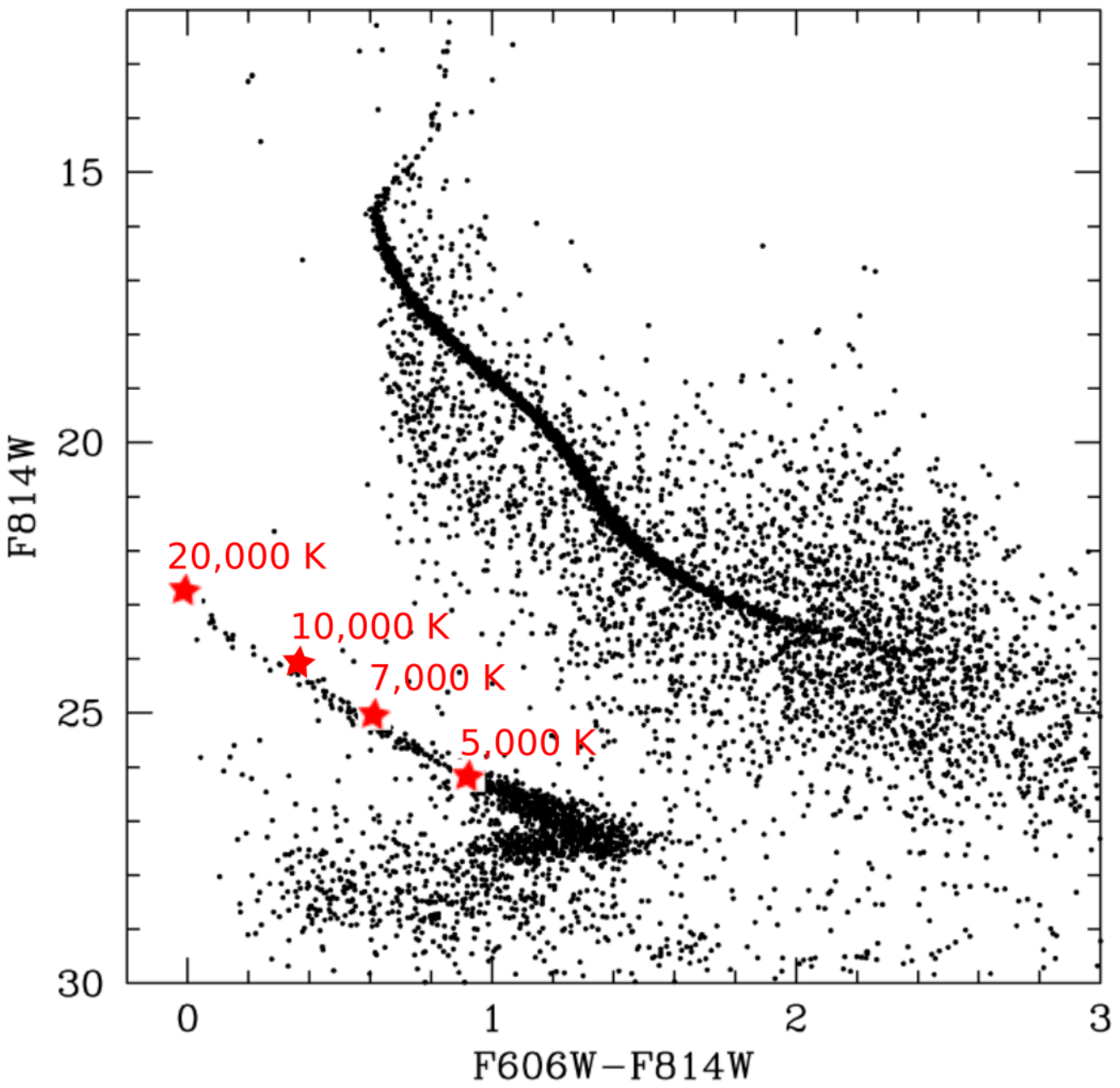}}
\caption
   {
CMD of a deep HST/ACS observation of a field in the globular cluster NGC 6397 \citep{2007ApJ...671..380H}.
The red stars and labels refer to blackbody models of white dwarfs  with effective temperatures of $5000$, $7000$, $10000$, and $20000$\,K.
   }
\label{Fig:Hansen2007}
\end{figure}

What could be the observational signatures of DM burning in WDs in a globublar cluster? In Fig.\,\ref{Fig:Hansen2007} we show a deep colour-magnitude diagram (CMD)  of a field in a globular cluster \citep{2007ApJ...671..380H}. The WD cooling sequence appears to the bottom left, at significantly fainter magnitudes and bluer colours than the main sequence. The data are deep enough to reach all the way to the bottom of the WD cooling sequence, which is truncated at $F814W\approx27.6$\,mag. We overplot the data points from four simple blackbody WD models over the data in the figure. As can be seen, the blackbody approximation works well for most of the cooling sequence, but at the coolest temperatures/faintest magnitudes, the cooling sequence turns toward bluer colours and deviates from what we would expect from simple blackbody cooling.

\begin{table*}
\caption{
Characteristics of the WD models, including their standard luminosities and the luminosities
due to DM burning in the scenarios without an IMBH at the center of $\omega-$Cen
($L_{\chi, \rm No\,BH}$, valid for $r<2.5\;$pc) and with an IMBH ($L_{\chi, \bullet}$) for WDs at distances of 2.5, 0.1 and 0.01 pc from it.
}
\begin{tabular}{|c|c|c|c|c|c|c|c|} 
\hline
Element & Mass & Radius & Luminosity & $T_{\rm eff}$ &
$L_{\chi, \rm No\,BH} = L_{\chi, \bullet} (r=2.5\, \rm pc)$ &
$L_{\chi, \bullet} (r=0.1\, \rm pc)$ & $L_{\chi, \bullet} (r=0.01\, \rm pc)$ \\
\ & ($M_{\odot}$) & ($R_{\odot}$) & ($L_{\odot}$) & (K) & ($L_{\odot}$) &
($L_{\odot}$) & ($L_{\odot}$) \\
\hline
\hline
He & 0.46 & 0.0160 & 0.00475 & 12000 & $4.79\times10^{-4}$ & $9.78\times10^{-4}$ & $2.47\times10^{-2}$ \\
C & 0.55 & 0.0149 & 0.162 & 30000 & $5.35\times10^{-4}$ & $1.09\times10^{-3}$ & $2.76\times10^{-2}$ \\
O & 0.55 & 0.0155 & 0.324 & 35000 & $5.55\times10^{-4}$ & $1.13\times10^{-3}$ & $2.87\times10^{-2}$ \\
\hline
He & 0.46 & 0.0152 & 0.000848 & 8000 & $4.55\times10^{-4}$ & $9.29\times10^{-4}$ & $2.35\times10^{-2}$ \\
C & 0.55 & 0.0145 & 0.0735 & 25000 & $5.19\times10^{-4}$ & $1.06\times10^{-3}$ & $2.68\times10^{-2}$ \\
O & 0.55 & 0.0147 & 0.141 & 30000 & $5.26\times10^{-4}$ & $1.07\times10^{-3}$ & $2.72\times10^{-2}$ \\
\hline
He & 0.46 & 0.0143 & 0.0000469 & 4000 & $4.28\times10^{-4}$ & $8.74\times10^{-4}$ & $2.21\times10^{-2}$ \\
C & 0.55 & 0.0141 & 0.0285 & 20000 & $5.05\times10^{-4}$ & $1.03\times10^{-3}$ & $2.61\times10^{-2}$ \\
O & 0.55 & 0.0143 & 0.0716 & 25000 & $5.12\times10^{-4}$ & $1.05\times10^{-3}$ & $2.64\times10^{-2}$ \\
\end{tabular}
\label{tab.all}
\end{table*}

\begin{figure*}
\resizebox{\hsize}{!}
{\includegraphics[scale=1,clip]{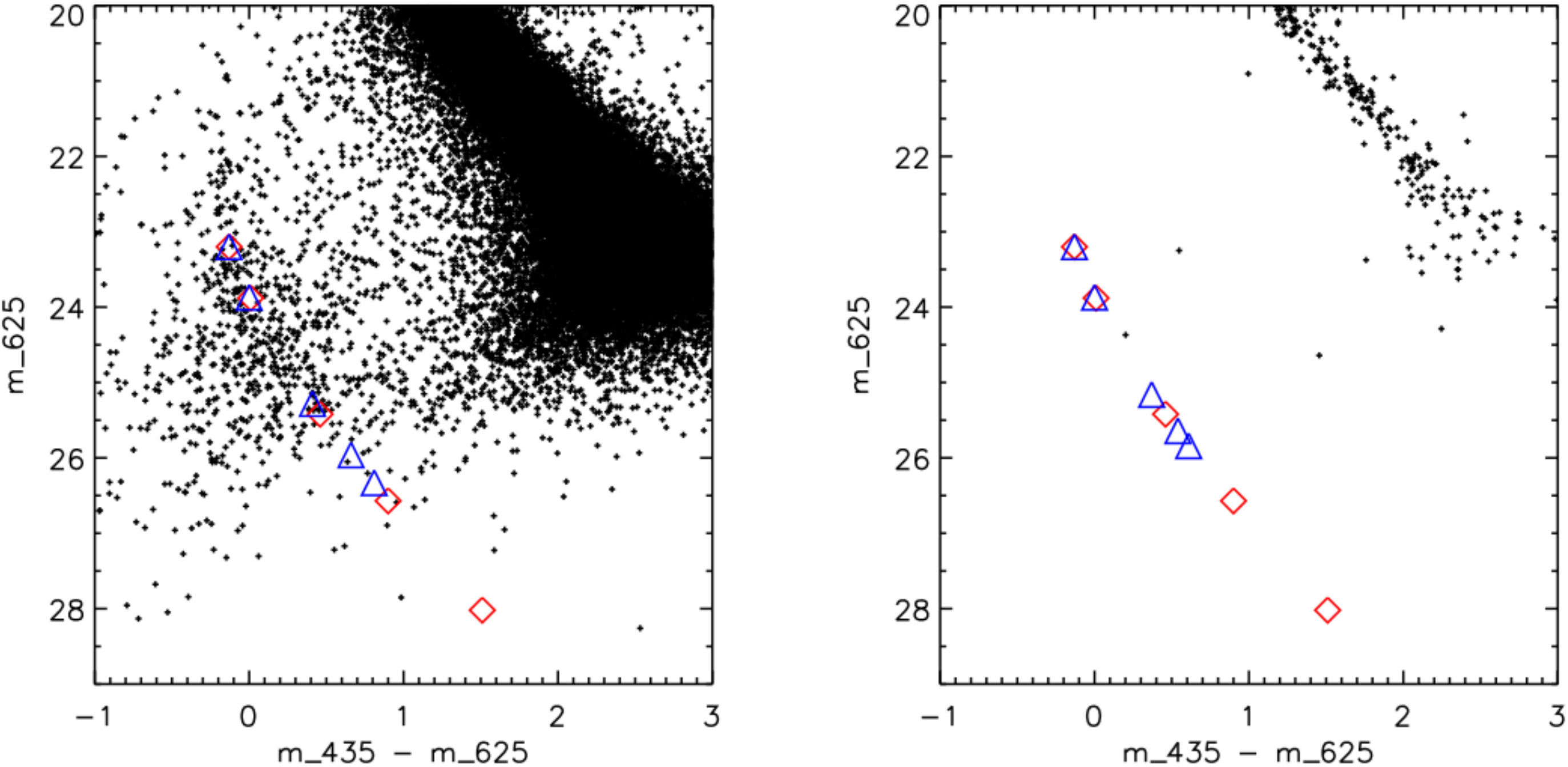}}
\caption
   {
CMDs of the stars detected within a projected radius of 2.5~pc (left) and 0.1~pc (right) from the center
of $\omega-$Cen \citep{AndersonvanderMarel2010}, with blackbody WD models overplotted.  Red diamonds correspond  to WDs at $T_{eff} = 30000, 20000, 10000, 7000$, and $5000$\,K. Blue triangles correspond to blackbody models of the same WDs with increased luminosity through DM burning in the core, i.e.\ within 2.5\,pc of the centre of the cluster (left: $T_{eff, DM} = 30000, 20100, 10600, 8350$, and $7500$\,K) and at $0.1$\,pc distance from  a hypothentical central IMBH (right: $T_{eff, DM} = 30050, 20150, 11100, 9300$, and $8700$\,K).
  }
\label{fig.OMCen_Anderson}
\end{figure*}

\cite{2000A&A...353..970P} calculated the mass-radius relationship for white dwarfs of different core compositions with the most common envelopes (hydrogen and helium) for a range of temperatures. We take 0.46 $M_{\odot}$ stars with He cores and 0.55 $M_{\odot}$ stars with C and O cores, since these were shown to be most common in $\omega$ -Cen by \citet{2013ApJ...769L..32B}.  Using these values, we then compute the luminosity from DM annihilation inside the WDs at different radial distances from the center of the $\omega-$Cen cluster, as explained in Sections~\ref{sec.rhochi} and \ref{sec.Lum}.   The results for different WD models and distances to the center of $\omega-$Cen are shown in Table~\ref{tab.all}.  We then add this luminosity to the luminosity of a WD that is not burning DM particles. Subsequently we calculate the total effective temperature of the WD while it is burning DM. Adding the effect of DM results simply in an increased $T_{\rm eff}$.  Observationally, this corresponds to the WD moving along the cooling sequence towards brighter magnitudes and bluer colours in a CMD. In particular, DM burning will impose a lower limit on the effective temperature of WDs, even if they are very old and had time to cool for almost a Hubble Time. Hence, {\it we predict  that DM burning will lead to a truncation of the cooling sequence at the faint end}.

To facilitate the following considerations, we note that, according to Table~\ref{tab.all}, the predicted DM luminosities of WDs differ by $\lesssim10\%$ for different WD compositions and temperatures and depend primarily on their distance from the cluster centre and on the presence of a central IMBH or not. Hence, we limit our computations to three cases: (1) DM burning within the cluster core and absence of an IMBH, (2) DM burning within the cluster core with an IMBH at a distance of 0.1\,pc from the latter, and (3) as (2) but at a distance of 0.01\,pc from the IMBH.  From Table~\ref{tab.all} we obtain for these cases mean DM luminosities of (1) $5.0\times10^{-4}$, $1.0\times10^{-3}$, and $2.6\times10^{-2}$ solar luminositis. We use a single mean radius of $0.015$ solar radii for all WD models.  We approximate the spectral energy distributions (SEDs) of the WDs by blackbodies.  In the blackbody-approximation, the luminosity/SED of the WDs is determined simply by their radius and effective temperature. The relation between these quantities and the effective temperature is given by the Stefan-Boltzmann law:
\begin{equation}
 L = 4 \pi R^2 \sigma T_{\rm eff}^4,
\end{equation}
where $L$ is the WD's luminosity, $R$ its radius, $T_{\rm eff}$ its effective
temperature, and  $\sigma \approx 5.6704\times 10^{-8}  Js^{-1}m^{-2}K^{-4}$.

For the properties and distribution of DM and stars  in $\omega$ -Cen that we assume in this paper, we obtain minimum effective Temperatures of $T_{eff}\approx7,000$\,K for WDs inside the core, i.e.\ within $2.5$\,pc of the centre of $\omega$ -Cen. If there exists an IMBH, then we obtain minimum values of  $T_{eff}\approx8,500$\,K for WDs within $0.1$\,pc and of  $T_{eff}\approx19,000$\,K for WDs within $0.01$\,pc of the black hole, inside the DM cusp.

In order to consider the observability of DM burning WDs we use HST ACS F435W
and F625W photometry of stars in $\omega-$Cen by \citet{AndersonvanderMarel2010}. We show the CMDs for regions within $2.5$\,pc and $0.1$\,pc projected distance from the centre of the cluster as adopted by \citet{AndersonvanderMarel2010} in Fig.\,\ref{fig.OMCen_Anderson}. We indicate the positions of WDs with different effective temperatures in these CMDs, as well as how they would be changed through DM burning. According to our assumptions, the WD cooling sequence should be truncated near $T_{eff} = 7500$\,K in the cluster core, independent of the presence of an IMBH. If there is an IMBH, then the cooling sequence would appear truncated around $T_{eff} = 8700$\,K for distances $<0.1$\,pc from the centre. Here, for the simplicity of argument, we neglect any projection effects. The latter would result in some lower temperature WDs from larger three-dimensional distances to appear below the truncation temperatures in the CMDs. We will not go through a detailed analysis of projection effects here because we are mainly interested in a zero order description of the potential effects of DM burning in WDs.

The CMDs of $\omega-$Cen show an almost complete absence of WDs at
$T_{eff}\lesssim10000$\,K and within $0.1$\,pc of the centre. This is, however,
merely an effect of incompleteness due to sensitivity and crowding. Not even the
sensitive, high-angular resolution observations made possible by the HST can
resolve the dense core of $\omega-$Cen sufficiently well to probe the WD
cooling sequence with high completeness down to low temperatures. Figure\,5 of
\citet{AndersonvanderMarel2010} indicates a completeness of only around 25\%
for stars of $mag_{F435W}\approx25$ within $0.35$\,pc of the cluster center.
Hence, incompleteness hinders us from drawing any meaningful conclusion on WDs
in the center of $\omega-$Cen. Crowding becomes less severe at greater
distances, but the situation does not change much, since the projected surface
density is rather flat in the core of $\omega-$Cen
\citep{AndersonvanderMarel2010}.  The completeness for stars of
$mag_{F438W}\approx25$ within $2.5$\,pc of the cluster center is still only
about 35\%. Therefore, although the CMD at the left panel of
Fig.~\ref{fig.OMCen_Anderson} may indicate a lack of WDs with
$T_{eff}\lesssim8000$\,K, this is not a reliable measurement. We also used
other observations \citep{2013ApJ...769L..32B}, but the situation remains
largely unchanged. With an estimated central density of
$5.6\times10^7\,M_{\odot}$\,pc$^{-3}$ \citep{art-Noyola08} the central regions
of $\omega-$Cen cannot be resolved down to faint magnitudes with high
completeness by any currently existing instrument.

The observational situation will not be much different for other globular clusters, at least the ones that are massive and dense enough to possibly hold an IMBH at their centres. A breakthrough can be expected with the advent of adaptive optics (AO) assisted imagers on the next generation of 30-40\,m-class telescopes. The near-infrared imager planned for the 39m ESO E-ELT, MICADO, for example, will provide an angular resolution of 6-12\,milli-arcseconds, depending on the observing wavelength. This will be almost five times better than what can be done with AO imagers on current 8m-class telescopes. Hence, confusion will be reduced by a factor of $>20$. Along with the high point-source sensitivity of the E-ELT this should make it possible to resolve the cores of nearby globular clusters and probe the WD cooling sequences down to their bottom.

Hence, the truncation of the WD cooling sequence at the cool end in dense globular clusters can serve as a future observational test that will set constraints on the properties of WIMP DM and, in the globular cluster cores, on the mass of a potential IMBH. If DM does indeed change WD luminosities in the predicted way, then by comparing the statistical properties of WDs in the central $\sim0.01-0.1$\,pc with their properties at distances of 1-2\,pc may even provide a conclusive answer to the presence of an IMBH, independent of stellar dynamical measurements.

\section{Conclusions}
\label{sec.conc}

We show that clusters are favorable environments to search for the DM effects
on stars and they could provide a tool to constrain the properties of DM. These clusters
are the only environments where we will be able to clearly observe stars in an
area rich in DM, unlike other obscured regions as the galactic center.

We found that DM burning in WDs would increase their effective temperature,
resulting in their moving up the cooling sequence. In addition, we show that
the minimum effective temperature of WDs due to DM burning strongly depends on
the presence of an IMBH at the center of the globular cluster. It will not be possible to
distinguish an individual DM burning WD from a normal one. However, we have
predicted a statistical signal if DM is present, particularly since WDs cool
faster at higher temperatures, meaning they should spend less time at the top
of the cooling sequence. If DM burning takes place, we expect a clear lack of
cool WDs near the cluster core and a distinct bunching of WDs at higher
temperatures. Deep, high angular resolution imaging, such as it could be provided by a futre 30-40m telescope will be needed to explore
the lack of an IMBH in
the center of $\omega-$Cen or, alternatively, it may prove that DM does not
have the large scattering cross-section on nucleons required to produce these
annihilation luminosities.

On the other hand, the observation of normal numbers of cool WDs in the centre
of $\omega-$Cen would be in disagreement with $\omega-$Cen harbouring an IMBH
and DM having the properties to explain the positive results in DAMA, CRESST
and CDMS experiments. Unfortunately, conclusive observational tests appear to
be currently out of reach.  The CMD from present HST observations of
$\omega-$Cen cannot provide any positive indication of DM burning due to the
lack of completeness of the observations in the reduced volume in the core of
$\omega-$Cen in which DM burning can be significant.  The high angular
resolution and sensitivity of future extremely large telescopes, such as the
GMT, TMT or E-ELT, will allow us to test our predictions by  searching for
truncated WD cooling sequences in globular clusters.

\section*{Acknowledgments}

PAS is thankful to Juan Barranco Monarca for the discussions at the AEI that
later led to the idea of this work, and to Luisa Seoane Rey for her
extraordinary support. We are indebted with Holger Baumgardt for providing us
with the data we used to study the distribution of WDs in radius, and with
M{\'a}rcio Catelan for comments on the observational aspects.  PAS is indebted
to the Universidad Cat{\'o}lica and the Department of Astronomy and the
Department of Physics of the University of Concepci{\'o}n for support during
his visit, and in particular to Julio Chanam{\'e}, Mike Fellhauer and Paulina
Assmann for their hospitality. JCa acknowledges support from the Alexander von
Humboldt Foundation. JCu acknowledges support from FONDECYT (1141175), Basal
(PFB0609) and the AEI for a visit to the institute.  PAS acknowledges the
hospitality of the Kavli Institute for Theoretical Physics. This research was
supported in part by the National Science Foundation under Grant No.  NSF
PHY11-25915.

\label{lastpage}


\begin{thebibliography}{49}
\expandafter\ifx\csname natexlab\endcsname\relax\def\natexlab#1{#1}\fi

\bibitem[{Abramowski {et~al.}(2011)}]{art-HESS11}
Abramowski A., {et~al.}, 2011, ApJ, 735, 12

\bibitem[{Agnese {et~al.}(2013)}]{Agnese:2013rvf}
Agnese R., {et~al.}, 2013, Phys. Rev. Lett., 111, 251301

\bibitem[{Akerib {et~al.}(2014)}]{2014PhRvL.112i1303A}
Akerib D.~S., {et~al.}, 2014, Phys. Rev. Lett., 112, 091303

\bibitem[{{Amaro-Seoane} {et~al.}(2004){Amaro-Seoane}, {Freitag}, \&
  {Spurzem}}]{ASEtAl04}
{Amaro-Seoane} P., {Freitag} M., {Spurzem} R., 2004, MNRAS

\bibitem[{{Anderson} \& {van der Marel}(2010)}]{AndersonvanderMarel2010}
{Anderson} J., {van der Marel} R.~P., 2010, ApJ, 710, 1032

\bibitem[{Angloher {et~al.}(2012)}]{2012EPJC...72.1971A}
Angloher G., {et~al.}, 2012, Eur.Phys.J.C, 72, 1971

\bibitem[{Bahcall \& Wolf(1976)}]{Bahcall:1976aa}
Bahcall J., Wolf R., 1976, ApJ, 209, 214

\bibitem[{{Bellini} {et~al.}(2013){Bellini}, {Anderson}, {Salaris}, {Cassisi},
  {Bedin}, {Piotto}, \& {Bergeron}}]{2013ApJ...769L..32B}
{Bellini} A., {Anderson} J., {Salaris} M., {Cassisi} S., {Bedin} L.~R.,
  {Piotto} G., {Bergeron} P., 2013, ApJ Lett., 769, L32

\bibitem[{Bernabei {et~al.}(2008)}]{Bernabei:2008yi}
Bernabei R., {et~al.}, 2008, Eur.Phys.J., C56, 333

\bibitem[{{Bertone} \& {Fairbairn}(2008)}]{2008PhRvD..77d3515B}
{Bertone} G., {Fairbairn} M., 2008, Ph.Rv.D, 77, 043515

\bibitem[{{Bertone} \& {Merritt}(2005)}]{BertoneMerritt05}
{Bertone} G., {Merritt} D., 2005, Physical Review D, 72, 103502

\bibitem[{{Blumenthal} {et~al.}(1986){Blumenthal}, {Faber}, {Flores}, \&
  {Primack}}]{art-Blumenthal86}
{Blumenthal} G.~R., {Faber} S.~M., {Flores} R., {Primack} J.~R., 1986, ApJ,
  301, 27

\bibitem[{{Bouquet} \& {Salati}(1989)}]{BouquetSalati89}
{Bouquet} A., {Salati} P., 1989, ApJ, 346, 284

\bibitem[{Burkert(1996)}]{Burkert:1995yz}
Burkert A., 1996, IAU Symp., 171, 175

\bibitem[{Casanellas \& Lopes(2009)}]{Casanellas:2009dp}
Casanellas J., Lopes I., 2009, ApJ, 705, 135

\bibitem[{{Fan} {et~al.}(2011){Fan}, {Yang}, \& {Chang}}]{2011PhRvD..84j3510F}
{Fan} Y.-Z., {Yang} R.-Z., {Chang} J., 2011, Ph.Rv.D, 84, 103510

\bibitem[{{Feng} {et~al.}(2012){Feng}, {Yuan}, {Yin}, {Bi}, \&
  {Li}}]{art-Feng12}
{Feng} L., {Yuan} Q., {Yin} P.-F., {Bi} X.-J., {Li} M., 2012, JCAP, 4, 30

\bibitem[{Gao {et~al.}(2004)Gao, White, Jenkins, Stoehr, \&
  Springel}]{Gao:2004au}
Gao L., White S.~D., Jenkins A., Stoehr F., Springel V., 2004, MNRAS, 355, 819

\bibitem[{Gnedin {et~al.}(2004)Gnedin, Kravtsov, Klypin, \&
  Nagai}]{Gnedin:2004cx}
Gnedin O.~Y., Kravtsov A.~V., Klypin A.~A., Nagai D., 2004, ApJ, 616, 16

\bibitem[{{Gnedin} \& {Primack}(2004{\natexlab{a}})}]{GnedinPrimack2004}
{Gnedin} O.~Y., {Primack} J.~R., 2004{\natexlab{a}}, Physical Review Letters,
  93, 061302

\bibitem[{{Gnedin} \& {Primack}(2004{\natexlab{b}})}]{2004PhRvL..93f1302G}
---, 2004{\natexlab{b}}, Phys. Rev. Lett., 93, 061302

\bibitem[{{Gondolo} \& {Silk}(1999)}]{GondoloSilk99}
{Gondolo} P., {Silk} J., 1999, Physical Review Letters, 83, 1719

\bibitem[{{Gould}(1987)}]{Gould1987}
{Gould} A., 1987, ApJ, 321, 571

\bibitem[{{Griffen} {et~al.}(2010){Griffen}, {Drinkwater}, {Thomas}, {Helly},
  \& {Pimbblet}}]{2010MNRAS.405..375G}
{Griffen} B.~F., {Drinkwater} M.~J., {Thomas} P.~A., {Helly} J.~C., {Pimbblet}
  K.~A., 2010, MNRAS, 405, 375

\bibitem[{{Hansen} {et~al.}(2007){Hansen}, {Anderson}, {Brewer}, {Dotter},
  {Fahlman}, {Hurley}, {Kalirai}, {King}, {Reitzel}, {Richer}, {Rich}, {Shara},
  \& {Stetson}}]{2007ApJ...671..380H}
{Hansen} B.~M.~S., {Anderson} J., {Brewer} J., {Dotter} A., {Fahlman} G.~G.,
  {Hurley} J., {Kalirai} J., {King} I., {Reitzel} D., {Richer} H.~B., {Rich}
  R.~M., {Shara} M.~M., {Stetson} P.~B., 2007, ApJ, 671, 380

\bibitem[{{Hooper} {et~al.}(2010){Hooper}, {Spolyar}, {Vallinotto}, \&
  {Gnedin}}]{Hooper:2010es}
{Hooper} D., {Spolyar} D., {Vallinotto} A., {Gnedin} N.~Y., 2010, Phys.Rev.D,
  81, 103531

\bibitem[{{Hurst} {et~al.}(2015){Hurst}, {Zentner}, {Natarajan}, \&
  {Badenes}}]{2014arXiv1410.3925H}
{Hurst} T.~J., {Zentner} A.~R., {Natarajan} A., {Badenes} C., 2015, Ph.Rv.D,
  91, 103514

\bibitem[{{Ibata} {et~al.}(2013){Ibata}, {Nipoti}, {Sollima}, {Bellazzini},
  {Chapman}, \& {Dalessandro}}]{Ibata:2012eq}
{Ibata} R., {Nipoti} C., {Sollima} A., {Bellazzini} M., {Chapman} S.~C.,
  {Dalessandro} E., 2013, MNRAS, 428, 3648

\bibitem[{{Jalali} {et~al.}(2012){Jalali}, {Baumgardt}, {Kissler-Patig},
  {Gebhardt}, {Noyola}, {L{\"u}tzgendorf}, \& {de Zeeuw}}]{JalaliEtAl2012}
{Jalali} B., {Baumgardt} H., {Kissler-Patig} M., {Gebhardt} K., {Noyola} E.,
  {L{\"u}tzgendorf} N., {de Zeeuw} P.~T., 2012, A\&A, 538, A19

\bibitem[{Mashchenko \& Sills(2005)}]{Mashchenko:2004hk}
Mashchenko S., Sills A., 2005, ApJ, 619, 258

\bibitem[{{McCullough} \& {Fairbairn}(2010)}]{McCullough:2010ai}
{McCullough} M., {Fairbairn} M., 2010, Phys.Rev.D, 81, 083520

\bibitem[{{McNamara} {et~al.}(2012){McNamara}, {Harrison}, {Baumgardt}, \&
  {Khalaj}}]{McNamaraEtAl2012}
{McNamara} B.~J., {Harrison} T.~E., {Baumgardt} H., {Khalaj} P., 2012, ApJ,
  745, 175

\bibitem[{Merritt(2004)}]{Merritt:2003qk}
Merritt D., 2004, Phys. Rev. Lett., 92, 201304

\bibitem[{{Merritt} {et~al.}(2007){Merritt}, {Harfst}, \&
  {Bertone}}]{Merritt:2006mt}
{Merritt} D., {Harfst} S., {Bertone} G., 2007, Phys. Rev. D, 75, 043517

\bibitem[{Moore {et~al.}(1998)Moore, Governato, Quinn, Stadel, \&
  Lake}]{Moore:1997sg}
Moore B., Governato F., Quinn T.~R., Stadel J., Lake G., 1998, ApJ, 499, L5

\bibitem[{{Moore} {et~al.}(1999){Moore}, {Quinn}, {Governato}, {Stadel}, \&
  {Lake}}]{MooreEtAl99}
{Moore} B., {Quinn} T., {Governato} F., {Stadel} J., {Lake} G., 1999, MNRAS,
  310, 1147

\bibitem[{{Moskalenko} \& {Wai}(2007)}]{MoskalenkoWai07}
{Moskalenko} I.~V., {Wai} L.~L., 2007, ApJ Lett., 659, L29

\bibitem[{Navarro {et~al.}(1996)Navarro, Frenk, \& White}]{Navarro:1995iw}
Navarro J.~F., Frenk C.~S., White S.~D., 1996, ApJ, 462, 563

\bibitem[{{Navarro} {et~al.}(1997){Navarro}, {Frenk}, \&
  {White}}]{NavarroFrenkWhite97}
{Navarro} J.~F., {Frenk} C.~S., {White} S.~D.~M., 1997, ApJ, 490, 493

\bibitem[{{Noyola} {et~al.}(2008){Noyola}, {Gebhardt}, \&
  {Bergmann}}]{art-Noyola08}
{Noyola} E., {Gebhardt} K., {Bergmann} M., 2008, ApJ, 676, 1008

\bibitem[{{Noyola} {et~al.}(2010){Noyola}, {Gebhardt}, {Kissler-Patig},
  {L{\"u}tzgendorf}, {Jalali}, {de Zeeuw}, \&
  {Baumgardt}}]{2010ApJ...719L..60N}
{Noyola} E., {Gebhardt} K., {Kissler-Patig} M., {L{\"u}tzgendorf} N., {Jalali}
  B., {de Zeeuw} P.~T., {Baumgardt} H., 2010, ApJ Lett., 719, L60

\bibitem[{{Panei} {et~al.}(2000){Panei}, {Althaus}, \&
  {Benvenuto}}]{2000A&A...353..970P}
{Panei} J.~A., {Althaus} L.~G., {Benvenuto} O.~G., 2000, A\&A, 353, 970

\bibitem[{{Peebles}(1972)}]{Peebles72}
{Peebles} P. J.~E., 1972, ApJ, 178, 371

\bibitem[{{Peebles}(1984)}]{art-Peebles1984}
{Peebles} P.~J.~E., 1984, ApJ, 277, 470

\bibitem[{{Salati} \& {Silk}(1989)}]{SalatiSilk89}
{Salati} P., {Silk} J., 1989, ApJ, 338, 24

\bibitem[{{Scott} {et~al.}(2009){Scott}, {Fairbairn}, \&
  {Edsj{\"o}}}]{Scott:2008ns}
{Scott} P., {Fairbairn} M., {Edsj{\"o}} J., 2009, MNRAS, 394, 82

\bibitem[{{Tsuchiya} {et~al.}(2003){Tsuchiya}, {Dinescu}, \&
  {Korchagin}}]{Tsuchiya:2003ev}
{Tsuchiya} T., {Dinescu} D.~I., {Korchagin} V.~I., 2003, ApJ Lett., 589, L29

\bibitem[{van~de Ven {et~al.}(2006)van~de Ven, van~den Bosch, Verolme, \&
  de~Zeeuw}]{vandeVen:2005hk}
van~de Ven G., van~den Bosch R., Verolme E., de~Zeeuw P., 2006,
  Astron.Astrophys., 445, 513

\bibitem[{{van der Marel} \& {Anderson}(2010)}]{2010ApJ...710.1063V}
{van der Marel} R.~P., {Anderson} J., 2010, ApJ, 710, 1063

\end{thebibliography}
\end{document}